\documentclass[aps,floats,prl,floatfix,twocolumn]{revtex4}
\usepackage{amsmath,amssymb,graphicx,xspace,units}

\graphicspath{{figures/}}

\newcommand{\vecl}{\ensuremath{\vec \ell}\xspace}

\begin{document}

\title{Statistically Locked-in Transport Through Periodic Potential Landscapes} 

\author{Ajay Gopinathan}

\altaffiliation{Present address:
  Dept.\ of Chemistry and Biochemistry, 
  University of California, Los Angeles, CA 90095, and
  Dept. of Physics, University of California,
  Santa Barbara, CA 93106.}

\author{David G.\ Grier}

\affiliation{Dept.\ of Physics and James Franck Institute,
  The University of Chicago,  Chicago, IL 60637}

\date{\today}

\begin{abstract}
  Classical particles driven through periodically modulated potential energy landscapes
  are predicted to follow a Devil's staircase hierarchy of commensurate trajectories 
  depending on the orientation of the driving force.  Recent experiments on colloidal
  spheres flowing through arrays of optical traps do indeed reveal such a hierarchy,
  but not with the predicted structure.  The microscopic trajectories, moreover,
  appear to be random, with commensurability emerging only in a statistical sense.
  We introduce an idealized model for periodically modulated transport in the
  presence of randomness that captures both the structure and statistics of
  such statistically locked-in states.
\end{abstract}

\maketitle

Objects driven through periodic potential energy surfaces face
a myriad of choices: either they follow the
driving force or they become entrained along any of the
commensurate directions through the landscape.
Variants of this problem appear in areas as diverse as
driven charge density waves \cite{brown84}, electronic energy states in two-dimensional
electron gases \cite{wiersig01}, atom migration on crystal surfaces \cite{pierrelouis01}, 
chemical kinetics, and flux flow in type-II superconductors \cite{reichhardt99}.
Quite recently, this problem was investigated \cite{korda02b} 
using a monolayer of colloidal spheres
in flowing water as a model system and a square array of
holographic optical tweezers \cite{hot}
to provide the periodic potential energy surface.
Depending on the array's orientation with respect to the driving
force, the spheres were observed to trace out a Devil's staircase
hierarchy of commensurate directions through the array, with
particular directions being preferentially selected over certain
ranges of orientations \cite{korda02b}.
Trajectories deflected by a
preference for commensurability are said to be kinetically locked in
to the lattice.

This anticipated result \cite{ott93} was accompanied by two surprises.
In the first place, not all kinetically locked-in states were 
centered on simple commensurate directions.
Still more surprisingly, particles' microscopic trajectories
in high-order locked-in states did not consist of
sequences of commensurate jumps,
but rather consisted of seemingly random lower-order
hops whose combination, nonetheless, were commensurate.
The appearance of statistical rather than deterministic commensurability
suggests an unexpected role for thermal randomness in structuring transport
through periodic potentials, and has been dubbed statistical lock-in
\cite{korda02b}.

\begin{figure}[tbhp]
  \centering
  \includegraphics[width=0.75\columnwidth]{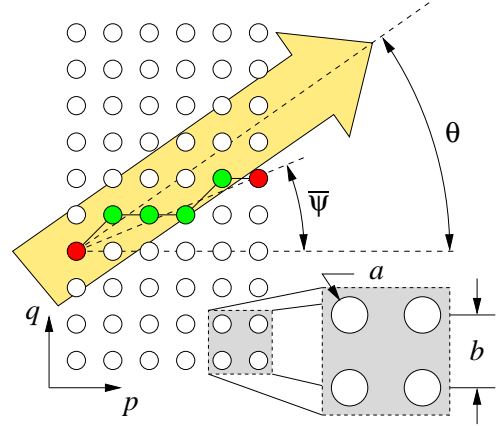}
  \caption{Schematic diagram of kinetically locked-in transport through
    a finite domain.  Potential wells of radius $a$ are arranged on
    a square grid of lattice constant $b$.  An applied force drives
    objects through the array at angle $\theta$, but, because of
    their interactions with the wells, the particles are most likely
    to follow a commensurate direction $\overline{\psi}(\theta)$ through the
    array.  Deflection by thermal forces or quenched random
    scattering centers causes the trajectories to deviate
    from commensurability, which is recovered only in a
    statistical sense.
  }
  \label{fig:schematic}
\end{figure}

This Letter presents a highly simplified model for statistically locked-in 
transport through mesoscopic potential energy landscapes
that nonetheless accounts for the emergence of combination jumps and their
statistical commensurability.
In particular this model reveals how the potential energy landscape's structure and extent
establish the discrete spectrum
of travel directions selected by biased random walkers.
While our discussion is directed toward the purely classical
behavior of flowing colloids, similar results should emerge for
biased quantum mechanical hopping through arrays of potential wells or
barriers.
Finally, we argue that the kinetically locked-in
state selected at a given orientation 
can depend sensitively on particle size.
Statistical lock-in therefore presents new
opportunities for continuously sorting
heterogeneous materials into multiple fractions simultaneously.

We first consider an array of optical traps, as shown in Fig.~\ref{fig:schematic}. 
The centers of 
the traps are on a square lattice with lattice constant $b$. 
The array is taken to extend indefinitely in the $y$ direction,
while its extent in the $x$ direction is $N$ lattice constants.
Rather than attempting to account for the detailed structure of a practical
trap array, we instead focus on the simplest model that 
captures the experimental phenomenology.
For this reason, we model the traps as circular regions of radius 
$a < b/2$ centered on grid points.
Once a particle enters a trap's domain, it is translated
directly to the center.
A particle passing outside the circle is 
assumed to be unaffected.
A ``trapped'' particle eventually escapes
and is carried along to its next encounter by the driving force. 
This is consistent with the observed
dynamics of colloidal particles flowing through holographic optical
trap arrays \cite{korda02b}.

In such a situation there are two distinct regimes we may consider.
The first is 
the ``ballistic'' regime where most particles pass through the lattice
without encountering traps.
This corresponds to the limit of small trap density, $\eta = a/b$.
For sufficiently large values of $\eta$, traps are packed closely enough together that
particles' trajectories consist mostly of jumps from one trap to another.
This we term the ``lattice gas'' regime.
We shall first derive the condition on $\eta$ for being in the lattice gas regime and consider only 
the case that meets this condition for the rest of the discussion.
Consider the array as shown in Fig.~\ref{fig:schematic}. 
Consider a line drawn from one of the lattice points on 
the left edge which we will take to be the origin. Let the line have a slope $m$. Now the shortest
 distance, $s$, between the line and a generic point on the lattice, represented by $(p,q)$, 
with $p \in \{1, \cdots ,N\}$ and $q \in \mathbb{Z}$, is given by
\begin{equation}
  s = \frac{b | q - pm | }{(1+m^{2})^{\frac{1}{2}}}
\end{equation}
We now wish to compute the largest possible value, $s_m$, of the distance from the 
line to the lattice 
point nearest to the line for all possible slopes. 
If the trap radius $a$ is larger than this distance 
then all straight line trajectories originating at a lattice point on the left side will necessarily
intersect a trap's region of influence.
Now the least upper bound on the minimum value of $| q - pm |$
can be shown to be $1/(N+1)$ \cite{steuard}. 
Using the smallest allowed value of $m$, which is the 
slope of the line through the origin and tangent to the circle centered at $(1,0)$, 
we estimate for an upper bound on $s_m$
\begin{equation}
  s_m \le \frac{1}{N+1} (b^2-a^2)^{\frac{1}{2}}
\end{equation}
Thus a sufficient (and necessary) condition that any straight line 
trajectory starting at the origin must
intersect a trap before exiting the lattice is $s_m = a$, or
\begin{equation}
  \eta > \frac{1}{\sqrt{(N+1)^2 + 1}}.
\label{eq:cond}
\end{equation}
For the remainder of the discussion we shall assume that this
condition is satisfied and that every trajectory involves a 
sequence of inter-trap jumps.

Every trajectory starting at a lattice point on the left side of the lattice and 
exiting at the right 
side can be decomposed into a discrete sequence of steps from one lattice point to another, 
followed by 
exit from the trap array.
For ease of modeling we treat the exit also as a step described by the lattice 
vector to the trap that would have captured the exiting particle had the trap array 
been extended to infinity in the $x$ direction.

\begin{figure}[t!]
  \centering
  \includegraphics[width=0.75\columnwidth]{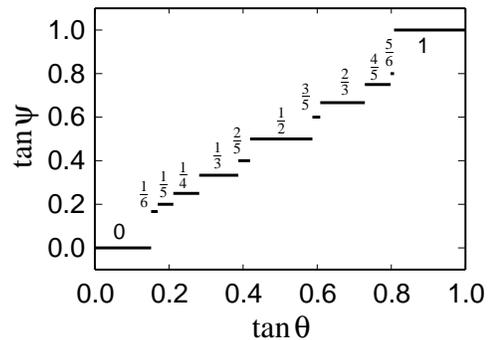}
  \caption{Direction of deterministically locked-in transport, $\tan \psi$, as a 
    function of driving direction $\tan \theta$.
    Results shown are for an array with trap density $\eta = a/b = 0.15$.}
  \label{fig:psi}
\end{figure}

In the physical system \cite{korda02b}, colloidal particles are driven
through an array of optical traps by a steady flow in the
surrounding medium oriented at angle $\theta$ to the 
trap array's [10] axis.
These particles are also subject to random thermal forces causing them to diffuse.
We will first consider the case where there is no diffusion.
In the absence of traps, the particles then would travel in straight lines
in the $\theta$ direction.
Within the trap array, however, a particle's displacement in each step is
determined by the position of the center of the next trap 
it encounters along its straight line trajectory. 
Thus, even in the absence of diffusion, the effective 
direction of motion can differ from the direction dictated by the flow.
The effective direction of 
motion will be described by an angle $\psi$ given by
$\tan \psi = \ell_y/\ell_x$ where 
$\vecl \equiv (\ell_x,\ell_y)$ is the lattice vector describing the step. 
The functional 
dependence of $\psi$ on the flow direction, $\theta$, is given by 
$\tan \psi = q/p$,  where $(p,q)$ 
are the smallest integers satisfying 
\begin{equation}
  |q - p \, \tan \theta| \leq \eta \, (1 + \tan \theta^2)^\frac{1}{2}
  \label{eq:psi}
\end{equation}
with $p \in \{1, \cdots ,N\}$.
Figure~\ref{fig:psi} shows the result.

Corresponding to a given value of $\theta$ there exists a unique value 
of $\psi$ and hence a lattice 
vector, $\vecl$, describing that step.
In general, a trajectory $T$ consists of a sequence of steps,
$T \equiv \{\vecl^1, \vecl^2, \cdots, \vecl^n\}$, where 
$\vecl^i$ is a lattice vector describing the $i$-th step of an $n$-step trajectory. Here 
$\vecl^i = (\ell^i_x,\ell^i_y)$, with $\ell^i_x \in \{1, \cdots, N\}$
and $\ell^i_y \in \mathbb{Z}$,
such that $\ell^i_x$ and $\ell^i_y$ are relatively prime, 
$\sum_{i=1}^n \ell^i_x \geq N$, and $\sum_{i=1}^{n-1} \ell^i_x < N$.
These conditions ensure that we do not count multiples of lattice vectors as distinct 
and that each trajectory terminates at the right side of the trap array. 
We can now define a \emph{mean} 
effective direction, $\overline{\psi}$, for the entire trajectory by
\begin{equation}
 \tan \overline{\psi}(T) = \frac{\sum_{i=1}^{n} \ell^i_y}{\sum_{i=1}^{n} \ell^i_x}
\end{equation}
This value characterizes the particles' overall transport through the array, 
and would be reflected in measurements of bulk transport properties, such as
a Hall voltage in the case of a periodically modulated two-dimensional
electron gas \cite{wiersig01}.
Without diffusion or some other randomizing process, all the steps in
every trajectory would be identical, with
$\overline{\psi} \equiv \psi$ for any orientation $\theta$.

We now consider the case in which diffusion or scattering randomizes
the particles' trajectories.
The natural assumption is that randomness would
smear out the noise-free transport characteristic in Fig.~\ref{fig:psi},
until, on average, the trajectories simply follow the driving direction,
$\overline \psi = \theta$.
In fact, far more interesting behavior results.

We model trajectory fluctuations by assuming that
a particle leaving a trap is equally likely to travel along any direction 
in a wedge of opening angle $2 \,\delta\theta$ centered on the flow direction. 
In terms of experimental parameters for colloidal transport, $\delta\theta$ 
depends on the temperature and the particles' mobility. 
The path a particle takes through the array 
is no longer deterministic, and there are many possible trajectories for any given 
driving direction. 

We first 
compute the probability of a step being a lattice vector $\vecl = (\ell_x,\ell_y)$ with 
$\ell_x \in \{1, \cdots, N\}$ and $\ell_y \in \mathbb{Z}$ 
such that $\ell_x$ and $\ell_y$ are relatively prime. 
This probability 
corresponds to the fraction of angles in the range 
$[\theta - \delta \theta, \theta + \delta \theta]$
whose effective direction $\psi$ corresponds to lattice vector $\vecl$. 
The probability is given by
\begin{equation}
  P_\ell(\vecl) = \frac{1}{2 \delta\theta} \int_{\theta-\delta\theta}^{\theta+\delta\theta} 
   \delta_{\frac{\ell_y}{\ell_x},\tan \psi(\theta^\prime)} \; d \theta^\prime,
\end{equation}
where $\delta_{j,k}$ is the Kronecker delta function.
If the condition in Eq.~(\ref{eq:cond}) 
is satisfied then there will only be a finite number of lattice
vectors with non-zero probability as defined above, and the sum of the probabilities will be
unity.
Consequently there is only a finite number of possible trajectories consisting of 
sequences of these lattice vectors. 
Let there be $M$ distinct possible trajectories given by the set $\{T_i\}$. 
Here $T_i = \{\vecl^1, \vecl^2, \cdots, \vecl^{n_i}\}$, where $n_i$ 
is the number of steps in trajectory $i$. 
The probability for a particle to take a particular trajectory 
$T_i$ is given by
\begin{equation}
  P_T(T_i) = \frac{\prod_{j=1}^{n_i} P_\ell(\vecl^j)}{%
    \sum_{i=1}^{M} \prod_{j=1}^{n_i} P_\ell(\vecl^j)}.
\end{equation}
The probability that the particle's trajectory will carry it in
direction $\overline{\psi_0}$, is then given by
\begin{equation}
  P(\overline{\psi_0}) = \sum_{i=1}^{M} P_T(T_i) \,
  \delta_{\tan \overline{\psi_0}, \tan \overline{\psi}(T_i)}.
\end{equation}
This distribution is nonzero for only a few discrete directions and thus is
better characterized by its most probable value, 
$\overline{\psi^\ast}$, with
$P(\overline{\psi^\ast}) \geq  P(\overline{\psi}) \; 
\forall \; \overline{\psi}$, rather than its mean.

\begin{figure}[t!]
  \centering
  \includegraphics[width=0.75\columnwidth]{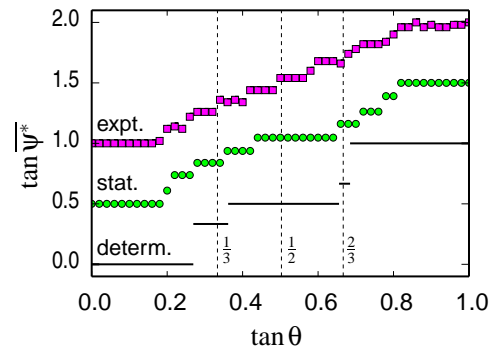}
  \caption{Statistically locked-in transport through an $N = 9$ lattice with
  $\eta = 0.26$ and $\delta \theta = 0.09~\unit{rad}$ (circles), compared with 
  deterministic transport (lines) and with experimental data
  from Fig.\ 3 of Ref.~\protect\cite{korda02b} (squares).  Data for statistically locked-in
  transport are offset by 0.5, and experimental data are offset by 1 for clarity.}
  \label{fig:expt}
\end{figure}

Figure~\ref{fig:expt} shows the results for an $N = 9$ lattice
with $\eta = 0.26$ when the
dispersal angle $\delta \theta$ that models disorder is increased from
the deterministic limit, $\delta \theta = 0$ to $\delta \theta = 0.09$.
These values were selected to mimic the experimental conditions in
Ref.~\cite{korda02b}.
Data from that study are plotted as squares in
Fig.~\ref{fig:expt} for comparison.
The calculated results demonstrate that
disorder does not necessarily wipe out the structure of
kinetically locked-in transport, but rather
reconfigures the pattern of plateaus.
In particular, not all statistically locked-in plateaus are
centered on commensurate directions.
This was one of the principal surprises to emerge from experimental studies
of colloid flowing through optical trap arrays \cite{korda02b}
and is a clear feature of the experimental data in Fig.~\ref{fig:expt}.
Now this restructuring can be explained as the emergence of high-order 
locked-in plateaus from a statistical sampling of low-order jumps.
Indeed, as the bottom trace in Fig.~\ref{fig:expt} demonstrates,
many of the commensurate directions corresponding to statistically
locked-in plateaus are not microscopically accessible at
$\eta = 0.26$.
These plateaus are absent, therefore, from the underlying pattern
of deterministically locked-in steps.

By contrast to the higher-order states,
the principal plateaus at $q/p = 0$ and $q/p = 1$ are both
microscopically and deterministically locked-in.
These correspond
to transport along the $[10]$ and $[11]$ lattice directions, respectively,
which cannot be decomposed into lower-index jumps.
This distinction also is clear in the experimental data \cite{korda02b}.
Under some conditions, other directions such as $[21]$ also can become
deterministically locked-in, particularly when both $\eta$ and
$\delta \theta$ are small.

The calculated transport characteristics do not agree
with the experimental data in all detail.
Qualitative differences, such as the jump in the experimental
data at $q/p = 1/2$ can be ascribed to the $10 \times 10$
structure of the experiment's optical trap array, which
differs from the extended geometry we considered here.
Flowing colloidal particles also are not drawn directly to 
optical traps' centers as they jump through a trap array,
but rather follow more subtle and
complicated trajectories.
Clearly, though these measurements more closely resemble our
model's predictions than they resemble the simple deterministic
spectrum for the same conditions.

The distribution of statistically locked-in states
changes dramatically with the size of the array.
Accessible states proliferate as $N$ increases, first
smearing out the spectrum of statistically locked-in states,
and then erasing even the deterministic plateaus.
In the large-$N$ limit and for $\delta \theta > 0$, the array has no overall
effect on transport:
$\overline{\psi^\ast} = \overline{\psi} = \theta$.
The hierarchically structured transport characteristics
discovered experimentally and explained here are features
of mesoscopic systems.

\begin{figure}[t!]
  \centering
  \includegraphics[width=0.75\columnwidth]{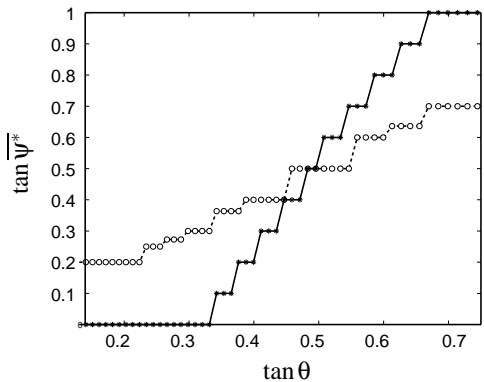}
  \caption{Dependence of the most probable transport direction,
    $\tan \overline{\psi^\ast}$, on
    driving direction $\tan \theta$ in the presence of
    random forcing.
    Circles show results for
    with $\eta = 0.15$, $\delta \theta = \pi/24$ and $N = 10$.
    The solid curve shows equivalent results for
    $\eta = 0.4$ and $\delta \theta = \pi/32$, and illustrates
    the extent to which qualitative features of the transport
    can change with the size of the flowing objects.}
  \label{fig:psimp}
\end{figure}

Being able to tune transport properties
by adjusting the number, distribution, and size of discrete potential
wells suggests a practical method for continuously separating 
heterogeneous fluid-borne samples into multiple fractions
simultaneously.
Consider, for example, objects driven through arrays of discrete
optical tweezers \cite{ashkin86}, each of which is formed by
a strongly focused beam of light.
Different objects passing through this optical intensity field
experience different potential energy landscapes depending
on their size, shape, and composition.
This effect already has been exploited to sort mixtures into
two fractions
with a single line of optical traps \cite{ladavac03}.
In terms of our model, the disk-like wells would appear to be larger
for larger or more strongly trapped objects.
Larger objects also would diffuse less vigorously, and so would
sample a simpler spectrum of statistically locked-in states.
Consequently, the same array of potential wells can have a substantially
different effect on different objects' transport.

Figure~\ref{fig:psimp} demonstrates that smaller particles
can be deflected into high-order statistically locked in states
under the same conditions that larger objects are deterministically
locked in to the $[10]$ direction.
The resulting angular separation is large enough that the two
resulting fractions could be collected, for example in microfluidic
channels.
Indeed, this figure makes clear that several fractions might be
continuously selected out of a mixed sample by a single static
array of optical traps.

We are grateful for extensive enlightening conversations with Tom Witten.
This work was supported primarily by the MRSEC program of the National Science Foundation
through Grant Number DMR-0213745 and in part by NSF Grants DMR-0304906 and
DBI-0233971.


\end{document}